\documentclass[compsoc,conference,a4paper,10pt,times]{IEEEtran}
\usepackage{amssymb}
\usepackage{float}
\usepackage{setspace}
\usepackage{enumitem}
\usepackage{cite}
\usepackage{url}
\usepackage{amsmath,amsfonts}
\usepackage{algorithmic}
\usepackage{graphicx}
\usepackage{textcomp}
\usepackage{bmpsize}
\usepackage{xcolor}
\usepackage{multicol}

\makeatletter
\newcommand{\linebreakand}{%
\end{@IEEEauthorhalign}
\hfill\mbox{}\par
\mbox{}\hfill\begin{@IEEEauthorhalign}
}
\makeatother

\newif\ifremarks 

\remarksfalse

\begin{document}

\title{Privacy and modern cars through a dual lens}

\author{\IEEEauthorblockN{1\textsuperscript{st} Giampaolo Bella}
\IEEEauthorblockA{\textit{Dipartimento di Matematica e Informatica} \\
\textit{Universit\`a degli Studi di Catania}\\
Catania, Italy \\
giamp@dmi.unict.it}
\and
\IEEEauthorblockN{2\textsuperscript{nd} Pietro Biondi}
\IEEEauthorblockA{\textit{Dipartimento di Matematica e Informatica} \\
\textit{Universit\`a degli Studi di Catania}\\
Catania, Italy \\
pietro.biondi@phd.unict.it}
\linebreakand %
\IEEEauthorblockN{3\textsuperscript{rd} Marco De Vincenzi}
\IEEEauthorblockA{\textit{Istituto di Informatica e Telematica} \\
\textit{Consiglio Nazionale delle Ricerche}\\
Pisa, Italy \\
marco.devincenzi@iit.cnr.it}
\and
\IEEEauthorblockN{4\textsuperscript{th} Giuseppe Tudisco}
\IEEEauthorblockA{\textit{Dipartimento di Matematica e Informatica} \\
\textit{Universit\`a degli Studi di Catania}\\
Catania, Italy \\
giuseppe.tudisco@studium.unict.it}
}

\maketitle

\begin{abstract}
Modern cars technologies are evolving quickly. They collect a variety of personal data and treat it on behalf of the car manufacturer to improve the drivers' experience.  
The precise terms of such a treatment are stated within the privacy policies accepted by the user when buying a car or through the infotainment system when it is first started.

This paper uses a double lens to assess people's privacy while they drive a car. The first approach is objective and studies the readability of privacy policies that comes with cars. We analyse the privacy policies of twelve car brands and apply well-known readability indices to evaluate the extent to which privacy policies are comprehensible by all drivers.
The second approach targets drivers' opinions to extrapolate their privacy concerns and trust perceptions. We design a questionnaire to collect the opinions of 88 participants and draw essential statistics about them. 
Our combined findings indicate that privacy is insufficiently understood at present as an issue deriving from driving a car, hence future technologies should be tailored to make people more aware of the issue and to enable them to express their preferences.

\end{abstract}


\section{Introduction}\label{sect:introduction}
Modern cars host highly developed technologies, such as infotainment systems and e-call boxes, routinely connected to the Internet. This increases the possible attack surface, and a number of examples of remotely hijacked cars exist. Cars may also collect drivers’ (or passengers’) personal data, hence privacy becomes a concern.

Intel estimates that a car can generate up to 4000 GB of data per day~\cite{intel2016}.
Thus, it is vitally important to understand and gather what types of (personal) data categories cars are collecting - and their manufacturers are treating - notably if these include special categories according to Regulation (EU) 2016/679, known as GDPR~\cite{gdpr}. In addition, it is important to understand whether and to what extent drivers are aware of what and how much data they disclose to car manufacturers and how this data is managed by them.
In fact, there would be limited use in addressing a problem that drivers did not feel.
Despite a few recent headlines on attacks to real cars~\cite{miller2014survey}, there is limited literature demonstrating how drivers feel about their privacy in their cars and what level of trust they pose e.g. in the interconnected infotainment systems that are becoming more and more common today, hence this objective. 

The goal of this paper is to analyse privacy issues in modern cars through a dual lens.
Through the first lens, we make an objective analysis of the availability and readability of the privacy policies the car makers provide to drivers as soon as they access to their service.
To do this, we analyse the documents  with respect to well-known readability indexes.
The second lens is a subjective one.
We constructed a ten-questions questionnaire in order to obtain the privacy concerns and trust perceptions of drivers with respect to modern cars.  To do this, we statistically analysed the responses of 88 drivers.
Our combined findings are that people's privacy concerns are rather low despite the moderate trust they generally express on their car's security and privacy measures, and this may be due to the opacity of the way privacy policies are written and presented. Therefore, it seems fair to claim that privacy is currently insufficiently understood as an issue deriving from driving a car.

\emph{The paper is structured as follows:}
Section~\ref{sec:arc} introduces the architecture of modern cars, Section~\ref{sec:privacyissues} explains in detail the privacy issues of this field, Section~\ref{sect:objectivepov} shows the study on privacy policies from an objective point of view, Section~\ref{sect:subjectivepov} describes the study from a subjective point of view, i.e. it analyses opinions from the point of view of privacy and trust of drivers with respect to modern cars,
Section~\ref{sec:RWandConcl} comments on related work and conclusions.

\section{Modern car architecture}\label{sec:arc}
Modern cars are computer on wheels. They have many different types of electronic control units (ECUs) on board that work together, thus enabling the complete operation of the car, they also provide and improve the usability and comfort of drivers and passengers.

In addition to control units, cars are composed of different sensors such as tyre pressure sensors, touch sensors that detects driver fatigue through grip, pulse or temperature sensors in the passenger compartment. Other sensors are installed above the roof of the car, such as radar or exterior cameras that take the car into the world of autonomous driving. In fact, modern cars tend to be increasingly connected to each other and to automotive infrastructures.

\subsection{Communication domains}

Modern cars appear to be increasingly complex and connected systems.
An increasing number of entities receive and transmit data through the connected vehicle ecosystem.
In particular, cars have several communication domains such as:
\begin{itemize}
    \item \textbf{Vehicle-to-Vehicle (V2V)} communication includes a wireless network where cars exchange messages with information about what they are doing. This data includes speed, position, direction of travel, braking and loss of stability. The aim of V2V communication is to prevent accidents by allowing passing vehicles to exchange position and speed data via an ad hoc network.
    
    \item  \textbf{Vehicle-to-Infrastructure (V2I)} communication is a communication model that allows vehicles to share information with the components that support a country's highway system. Such components include cameras, traffic lights, lane markers. Therefore, sensors can capture infrastructure data and provide travellers with real-time alerts on issues such as road conditions, traffic congestion and parking availability.
    
    \item \textbf{Intra-Vehicle communication (IV)} is the communication model in which the ECUs and sensors communicate with each other and exchange messages about the car's status. All this allows a car to function properly.
    
    \item \textbf{User-to-Vehicle communication (U2V)} concerns the inclusion of the smartphone, its data and functionalities (e.g. to allow easy hands-free dialling of the address book) via Bluetooth, Wi-Fi or USB mirroring.
    
    \item \textbf{Car maker and third-party services} need to collect data. The transmission of data to these services takes place through the use of a SIM card inside the car that provides connectivity to the car. In this scenario, the car transmits data both to the car maker's servers and to third-party services (e.g. for software updates or with satellites for geolocation), which are often accepted during the acceptance of privacy policies.
    
    \item \textbf{Emergency services} is one of the on board services. Thanks to the car's built-in SIM, the car can call the emergency services in case of a rescue need, such as a brutal impact that caused the airbags to deploy.

\end{itemize}

\begin{figure}[ht]
    \centering
    \includegraphics[scale=0.24]{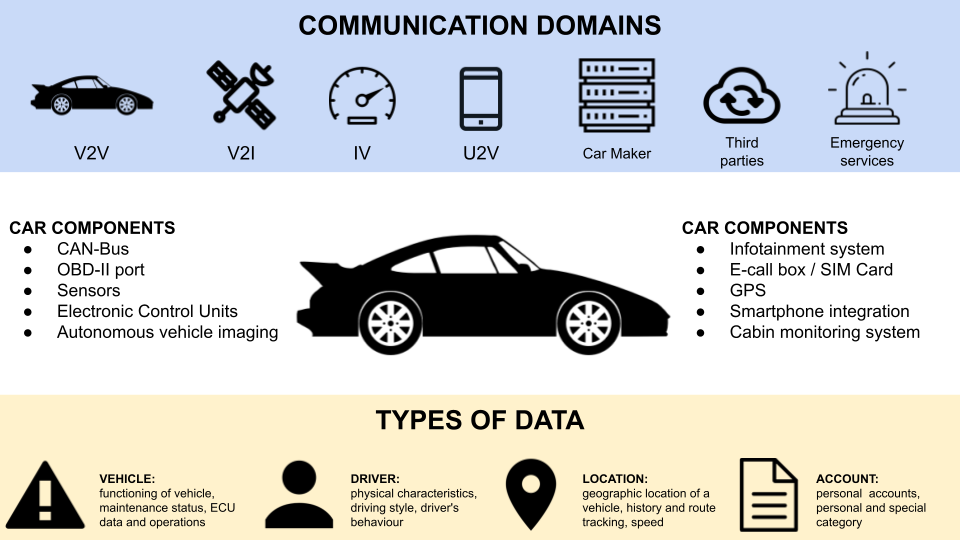}
    \caption{Infographics about data receivers, data type and car components.}
    \label{fig:infografica}
\end{figure}

The information about a modern car, the data it handles and transmits, and the components from which it is made can be summarised in the Figure~\ref{fig:infografica}.

\subsection{Data treatments}

The car architectures are based on data and the latter are transmitted to the infrastructures via the communication domains. All these data can be classified into several types:

\begin{itemize}
    \item \textbf{Vehicle Data} are all data concerning vehicle operation, maintenance status, mileage, tyre consumption, data from sensors. 
    
    \item \textbf{Driver Data} relates to driver profiling.  This can be done by sampling the physical characteristics or habits of the driver, such as, driving style of the vehicle, seat belt use, braking habits. 
    
    \item \textbf{Location Data} includes data, such as the geographical position of a vehicle, route history and tracking, speed, direction of travel.
    
    \item \textbf{Account Data} contains all data relating to the driver's personal accounts.

\end{itemize}

Processing data provides a number of benefits to drivers. Vehicle sensors and internal components can produce data for diagnostic purposes to check the health of the vehicle. Predicting and preventing component failures can effectively improve driver safety. In addition, processing sensor data allows technicians to conduct remote diagnostics and, by analysing historical data, repair costs can be lower.
The collection of data can also be useful in emergency situations. In fact, all cars sold in Europe since April 2018 must implement an emergency-call system, which is called eCall, as a measure to reduce fatalities caused by road accidents~\cite{eCall}. The eCall system is an electronic device that automatically alerts the emergency services in the event of a car accident. This system makes rescue operations faster and more effective and helps to save lives. When a severe impact is detected via the motion sensors, using the GPS and the vehicle's built-in SIM card for communication, data such as the exact position of the crashed car, direction of travel and type of impact are sent to the emergency services.

External service providers can use data to offer benefits to drivers, including economic benefits. These include usage-based car insurance, also known as pay-as-you-drive insurance. Insurers monitor the driving habits through a telematics device installed on the vehicle. By analysing data such as mileage, hard braking, rapid accelerations and cornering, the insurer is able to adapt the premium amount based on the driver's behaviour. The choice of a usage-based insurance over a traditional one has many potential benefits to the drivers. Thanks to the telematics device, drivers can monitor their driving habits and may get motivated to correct them because good drivers benefit from discounted premiums~\cite{UBI}. It is also easier to investigate in car accidents by having access to the events that took place before the accident occurred.

\section{Privacy Issues}\label{sec:privacyissues}

Nowadays, when we think of privacy breaches, we immediately consider smartphone applications or online services that leak users' personal information. 

In addition to smartphone and personal devices, also car cybersecurity is closely related to driver privacy, as modern cars collect a wide variety of personal data from drivers as a fundamental basis for the user experience. 
However, it is not easy to understand how cars exploit this data and, in particular, to what extent drivers understand the relevant implications, including the fact that car manufacturers are required to handle personal data in full compliance with relevant regulations. If car manufacturers collect data from a user in the European Union, this data is subject to the GDPR. Therefore, regardless of the geographical location of the servers that actually store the data, the company must ensure an adequate level of protection and privacy as required by the legislation.

The infotainment system also interacts directly with the driver and passengers by providing them with various functions, but these require data from users in order to properly work.
By synchronising personal devices with the infotainment system, users are able to stream multimedia files, make calls using the car's speakers, read e-mails and send text messages. They can also surf the web using the system's browser. The infotainment system can provide directions and traffic information using the vehicle's real-time location. Compared to sensors, this is the main component that makes use of the personal data of its users.
In addition, there is a variety of data concerning the personal data of drivers and passengers, including music preferences, favourite places, contact list, text messages and call logs.
In particular, we note that this group may include ``special categories of personal data'', i.e. data on health, political and religious orientation, biometric data, ethnic origin, as defined by Article 9 of the GDPR. For example, the car manufacturer may infer certain health conditions of the driver by intersecting historical data from seat weight sensors with the chosen interior temperature, seat back adjustment, car position or heart rate via sensors on the car steering~\cite{heartSteering}. Remarkably, this data set can also include the driver's financial data to pay for fuel and parking~\cite{car_payments}.

Furthermore, through vehicle geolocation we may be able to obtain information that is repeated over time and thus derive a possible habit, hobby or routine of a driver. For example, we may notice that every Sunday at a certain time the car is near a church, so we could deduce that the driver goes to church every Sunday and through the information obtained in an open way from the network we are able to deduce also the information regarding the type of religion that the driver professes.
Another example is recognising the driver through the seat sensor, which automatically adjusts the previously stored position based on weight.
These technologies can also monitor eye movement to detect a driver's attention in order to determine whether a driver is falling asleep at the wheel~\cite{personalDatainyourcar}. It is clear that all these sensors allow for an almost perfect identification of a driver and this increases the risk to the driver's privacy.

Finally, we can conclude that nowadays, more and more experiments show that cars can also manage the driver's personal data, often transmitted in the infotainment network, such as driving style~\cite{driverFingerprint}, location history \cite{bernardi_driver_2018} and also more general data such as cabin preferences, music preferences, and credit card details~\cite{edpbprivacy}.
However, the literature shows how the infotainment system can provide an entry point for attackers. Some examples involve exploiting vulnerabilities in the infotainment of a General Motors car to steal data from the remote system~\cite{GM2015}. Additionally, a few years ago, researchers discovered a number of vulnerabilities that, when combined, allowed them to remotely hack a Tesla Model S~\cite{Tesla2016}.

\section{Through the first lens: objective point of view}\label{sect:objectivepov}
Car makers collect and manage drivers' data. As soon as a driver uses one of the car maker's service, she has to provide the consensus by signing a \emph{privacy policy}, where each car maker should declare the type of data that it collects from the driver and from the car. In particular, following Article 12 of the GDPR, the controller of the data, in this case the car company, has to provide information to the user \textit {\lq\lq in a concise, transparent, intelligible and easily accessible form, using clear and plain language\rq\rq}~\cite{gdpr}. Our main target is to understand how easy it is for people to understand their data privacy in cars and the compliance with regulation like GDPR. In this section we present our analysis of privacy policies of twelve car makers and we consider a set of readability indexes evaluating the policy with respect to each of them.

\subsection{Privacy policy collection}
As a first step, we consider twelve car makers company: the top ten most famous car companies in Europe according to~\cite{bekker} (Audi, BMW, Ford, Mercedes, Opel, Peugeot, Renault, Skoda, Toyota, Volkswagen), Tesla, since it is the car company that most equips its cars with advanced technology, and KIA since we have a vehicle that we use for our cyber-security activity research. All the car manufacturers analysed in this work are shown in Table~\ref{tab:metric}.

Then, we download the privacy policy by using two different channels: during the installation of the respective app or from the company website and contacting the customer service of the car maker. The aim is being sure to analyse the most recent privacy policy of the considered car maker. 
Hence, we verify the correctness of the selected privacy policies by comparing the two of them and considering the one retrieved contacting the customer services and those one obtaining by using each car maker's app.
The customer services were contacted using the social network \lq\lq Facebook\rq\rq.
However, we received the required documents in six cases over twelve. Hence, it is necessary to contact via email or by phone another call center. 
The result of this double check is that the privacy policies downloaded from the app, are the same sent from the customer services. So, we can confirm that for a user it is quite easy to access and obtain the most updated privacy policy document. 
However, it is sometimes quite difficult to download the document and, without any particular translation, it is only possible to read it online, but this does not affect the possibility of being informed.

\subsection{Policy readability analysis}
Once we have established that the driver can access the privacy policies quite easily, we analysed whether the content is easy to be understood.
A text analysis on privacy policy documents (written in English) is performed by using \lq\lq textstat\rq\rq, a Python library to calculate statistics from text, that allows also to compute readability indexes \cite{textstat}. From the analysis, several metrics, such as, the \textit{number of words} or the \textit{number of sentences} are extracted. 
We consider also other metrics. In fact, one of the most important features of a text, especially of a privacy policy, is the \textit{readability}, that is the document quality of being easy and enjoyable to read \cite{cambridge1}.
By the word \lq\lq easy\rq\rq we mean that the policy must be in an easily accessible form with clear and simple language.

To understand if a text is readable, several indexes can be calculated. In this specific case, using the same Python library \textit{textstat}, we calculate the \textbf{Coleman-Liau} index, the \textbf{SMOG} index, the \textbf{Automated Readability} Index and the \textbf{Flesch Reading Ease} Index. The first three indexes use the U.S. school grade to label a text as \lq \lq difficult\rq\rq~or \lq \lq easy\rq\rq~to read. The U.S. schools have different grades, starting from 1 to 17 or more, that is graduated level. The 13th grade or above is considered university level. Table~\ref{table:conversion} shows an approximate comparison between the index scores and the US education level~\cite{conversion}.

\begin{table}[hbt!]
\centering
\caption{Table comparing Scores and Education Levels \cite{conversion}.}\label{table:conversion}
\renewcommand{\arraystretch}{1.3}
\scalebox{0.8}{
\begin{tabular}{|c|c|}
\hline

\textbf{Score/Grade} & 
\textbf{Education Level}\\ \hline

1-4 & Elementary School \\ \hline
5-8 & Middle School \\ \hline
9-12 & High School\\ \hline
13-16 & Undergraduate  \\ \hline
17+ & Graduate \\ \hline

\end{tabular}}
\end{table}

The \textbf{Coleman-Liau} Index (CLI) \cite{coleman}
depends on the complexity of the words, measured from the number of letters, and the complexity of the sentences.

The \textbf{SMOG} Index \cite{mclaurin}, acronym for \lq\lq Simple Measure of Gobbledygook\rq\rq~uses the polysyllables (words of 3 or more syllables) in a certain number of sentences (at least 30).

Despite the similarity with the previous indexes, the \textbf{Automated Readability} Index (ARI) \cite{ARI} takes into consideration also the number of characters, in addition to the number of words and sentences.

The fourth is the \textbf{Flesch Reading Ease} Index (FREI) \cite{flesch}, that differs from all the three previous indexes because it outputs a score instead of the school grade. 
The score ranges from 0 to 100 and the lower value indicates a text extremely difficult to read. It 
uses the number of words, sentences and also the number of syllables.

To combine the results obtained by the previous indexes into a single metric, the last column of Table \ref{tab:metric} presents a \emph{General Index of Difficulty of Readability (GIDR)} that is calculated by combining the previous four indexes: the lowest value ``{}0'' indicates the most readable privacy policy, while the value ``{}100'' the most difficult among the selected documents.
GIDR is formulated like Equation~\ref{eq:GIDR} and it is built starting from a harmonic mean. To merge the obtained indexes, we combined with a simple average the three indexes (CLI, ARI, SMOG) that have the same comparable scale (the US school grade). Then, we merged this mean with the FREI values. Due to the different scales of the mean and the FREI values, we chose the harmonic mean because it allows us to find division relationships between fractions without worrying about common denominators. The obtained value was normalised in a range between 0 and 100 using the Equation \ref{eq:GIDR_norm}. In Table \ref{tab:metric}, we summarise all findings obtained by analysing the twelve privacy policies.

\begin{equation}\label{eq:GIDR}
GIDR =  ({\frac{1}{\frac{CLI + SMOG + ARI}{3}}+\frac{1}{100-FREI}})^-1
\end{equation}

\begin{equation}\label{eq:GIDR_norm}
GIDR{_n} =  100 \times \frac{GIDR - min (GIDR)}{max(GIDR) - min (GIDR)}
\end{equation}

\begin{table}[t]
\caption{Metrics established for each privacy policy ordered according to the GIDR}\label{tab:metric}
\centering
\renewcommand{\arraystretch}{1.3}
\resizebox{\columnwidth}{!}{%
\begin{tabular}{|c|c|c|c|c|c|c|c|}
\hline
\multicolumn{8}{|c|}{\textbf{Privacy Policies Metrics}} \\ \hline
\textbf{Company} & \textbf{Number of Words} & \textbf{Number of Sentences} & \textbf{CLI} & \textbf{SMOG}& \textbf{ARI}& \textbf{FREI}& \textbf{GIDR}\\ \hline
Ford & 9744 & 1327 & 8.7 &  10.0 & 5.1	 & 58.3  & 0.0 \\ \hline
Peugeot & 2151 & 437 & 9.0 & 9.0 & 6.0 & 47.7  & 3.6\\ \hline
Kia & 22043 & 3096 & 9.6 & 10.4 & 5.8 & 49.8  & 10.7\\ \hline
Skoda & 5831 & 860 & 9.9 &  9.9 & 6.1 & 49.7  & 12.5\\ \hline
Mercedes & 8591 & 1387 & 10.0 &  10.0 & 6.0 & 44.0  & 14.3\\ \hline
Opel & 2438 & 323 & 11.0 & 10.0 & 7.0 & 46.6  & 21.4\\ \hline
Audi & 13661 & 1410 & 10.4 & 11.1 & 6.8 & 49.4  & 23.2\\ \hline
BMW & 991 & 119 & 11.1 & 10.5 & 7.1 & 49.1   & 25.0 \\ \hline
Tesla & 13224 & 1453 & 10.8 & 11.1 & 7.0 & 49.5  & 25.0\\ \hline
Volkswagen & 11742 & 1206 & 12.2 &  11.8 & 8.3 & 42.1  & 42.9\\ \hline
Toyota & 3279 & 263 & 11.7 &  13.2 & 8.8 & 43.6  & 48.2 \\ \hline
Renault & 2568 & 94 & 12.7 &  17.3 & 15.9 & 37.3  & 100.0\\ \hline
\end{tabular}
}
\end{table}

As a result, Table~\ref{tab:metric} shows that the number of words is a relative parameter to establish whether a policy is readable or not. It is not possible to identify any particular cluster, because, for instance, car makers belonging to the same group, like Volkswagen-Audi-Skoda or Peugeot-Opel, have different GIDR values. Instead, it is possible to note the concentration of the most car makers (9 over 12) in the first quartile of the GIDR value distribution, meaning a similar readability. However, inside this last set, Ford has a significant lower value than the others, resulting the most readable text. This difference seems to be determined by the FREI value, that, with respect to the others three readability indexes, is the only considering the number of syllables. This situation means that Ford could be the easiest readable text, because it uses shorter words with respect to others, and especially with respect to Renault.  

As far as we know, the three indexes, CLI, SMOG and ARI are considered three relevant indexes to evaluate text readability. They have been used since the 1960s/70s to define the scholastic level necessary for the comprehension of a text, starting from different values and coefficients as defined in the respective equations.
In our analysis, the SMOG and CLI indexes show values close to 11, that correspond to one of the last years of high school.
The ARI index, probably due to its equation, has slightly lower values, on average at a middle/high school level, but directly proportional and in line with the other two indexes, SMOG and CLI.

To summarise our analysis, it can be stated that the reading and the comprehension of the twelve privacy policy documents requires, on average, a high level of education equal to the last years of high school or the first years of university to be comprehensible in every part. This situation, for example in Europe, can affect the possibility of people to be informed in a concise, transparent and easily form as stated by GDPR \cite{gdpr}. In fact, according to \cite{eustat}, in 2020 in Europe only 35.9\% of people aged 25–54 has tertiary educational attainment and 21.9\% among people aged 54-74.

\section{Through the second lens: subjective point of view}\label{sect:subjectivepov}

In this section, conversely with respect to the previous one,  we show a subjective analysis of drivers' privacy concerns and trust perceptions. The following sections show the questionnaire we used for the study and the analysis of the findings obtained.
\subsection{Questionnaire approach and structure}
To explore the issue of privacy in modern cars, we decided to conduct a study of drivers using a questionnaire to ask about their concerns about privacy and perceptions of trust in modern cars.
Specifically, the questionnaire is divided into three parts. The first part deals with basic information about the respondents, including essential demographic data, including for the purpose of confirming that they are over 18 years old, and a check on the reliability of their answers.
The second part deals with respondents' privacy concerns and the third part with their perceptions of trust, so these two parts contain the key questions for our study. 

A 7-point balanced Likert scale is used for most of the questions in the questionnaire.  This is the most commonly used approach to scale responses in a research survey. 
Moreover, the range captures the intensity of their feelings for a given item. As such, likert scales have found application in psychology and social sciences, statistics, business and marketing~\cite{likertscale}.

In more detail, in Appendix~\ref{app:questionnaire}, we show the core questions of the questionnaire and the type of answers allowed.
In particular, the questionnaire begins with a question (Q0) that asks participants how many hours a week they spend driving their vehicle. This information can be interesting because an individual who spends a lot of time driving is likely to have a greater knowledge of the features provided by their car compared to an occasional driver.
The next question (Q1) asks participants to evaluate their knowledge about modern cars.
Then question Q2 asks respondents whether or not they agree that modern vehicles are similar and comparable to modern computers.
Question Q3 introduces the part related to the collection and processing of personal data. The question asks participants to select the kind of data they think modern cars collect.
Then question Q4 asks participants whether or not they agree that personal data collected by a modern car about its driver is necessary for the full functioning of the car. This question allows us to understand if drivers believe that, in order to be able to use all the features provided by their vehicle without any sort of limitation, they need to provide their personal data.
Question Q5 asks whether or not respondents think it is necessary to transmit the personal data collected over the Internet. In relation to the previous question, participants may agree to provide their personal data to obtain additional features (e.g. statistics about driving style and vehicle usage).
Question Q6 asks participants whether or not they agree that a modern vehicle safeguards its driver's life. This question introduces the survey part about the trust that drivers pose in their car.
Question Q7 asks respondents whether or not they agree that a modern car protects its driver's personal data better than it safeguards its driver's life.
Then with question Q8 participants are asked if they agree that a modern car processes the personal data it collects about its driver in a legitimate way that is consistent with the relevant regulations (e.g. \textit{GDPR}).
Question Q9 asks participants whether they agree that a modern car carries out a systematic and extensive evaluation of the personal data it collects about its driver on the basis of automated processing in order to evaluate personal aspects. In fact, according to the current legislation (art. 22 of the GDPR) these processes must be properly declared and explicit consent is required for the proposed purposes.
In addition, (art. 32 of) the GDPR requires the use of adequate security measures to protect the rights and freedoms of data subjects. We ask participants about it with the last question (Q10) where they are asked whether or not they agree that the personal data a modern car collects about its driver is protected by suitable technology when the car transmits data over the Internet. 

\subsection{Analysis of findings}

We have submitted the questionnaire to friends and colleagues to refine the methodology of the questions and the analysis of the findings, and we also think it could be a valid sample to be translated into a larger sample through crowdsourcing.
The result discussed in this chapter is based on a sample of 88 people who responded to the survey described above, moreover, we anticipate that these findings are very promising.

Starting with the first question, Q0 asks participants how many hours per week they spend driving their vehicle. The answers are shown in Table~\ref{tab:question0}. It turns out that 80\% of the participants do not drive more than 9 hours per week, this is probably due to the fact that many people have reduced their mobility by car because of the pandemic.

Then participants were asked to evaluate their knowledge about modern cars.
Considering the values in Table~\ref{tab:question1}, it can be affirmed that the interviewed sample considers itself knowledgeable about modern cars. Just a minority of participants (about 13\%) think they are not sufficiently knowledgeable about modern cars. Moreover, 17\% of participants think they have average knowledge while the rest of them (70\%) are quite confident about their knowledge.

\begin{table}[ht]
\parbox{.33\linewidth}{
\centering
\caption{Answers to the Q0 }
\label{tab:question0}
\scalebox{0.7}{
\begin{tabular}{l|c|}
    \cline{2-2}
    \multicolumn{1}{l|}{}                      & \textbf{\begin{tabular}[c]{@{}c@{}}Q0\end{tabular}} \\ \hline
    \multicolumn{1}{|c|}{\textbf{3-6 hours}}   & 38                                                            \\ \hline
    \multicolumn{1}{|c|}{\textbf{7-9 hours}}   & 32                                                            \\ \hline
    \multicolumn{1}{|c|}{\textbf{10-12 hours}} & 13                                                            \\ \hline
    \multicolumn{1}{|c|}{\textbf{13-15 hours}} & 1                                                             \\ \hline
    \multicolumn{1}{|c|}{\textbf{16-20 hours}} & 1                                                             \\ \hline
    \multicolumn{1}{|c|}{\textbf{21+ hours}}   & 3                                                             \\ \hline
\end{tabular}}
}
\hfill
\parbox{.65\linewidth}{
\centering
\caption{Answers to the Q1}
\label{tab:question1}
\scalebox{0.7}{
\begin{tabular}{l|c|}
\cline{2-2}
& \textbf{\begin{tabular}[c]{@{}c@{}}Q1\end{tabular}} \\ \hline
\multicolumn{1}{|l|}{\textbf{Knowledgeable about modern cars}}     & 70\%                                                            \\ \hline
\multicolumn{1}{|l|}{\textbf{Average knowledge}}                   & 17\%                                                            \\ \hline
\multicolumn{1}{|l|}{\textbf{Not knowledgeable about modern cars}} & 13\%                                                          \\ \hline
\end{tabular}}
}

\parbox{\linewidth}{
\centering
\caption{Answers to the Q3}
\label{tab:question3}
\scalebox{0.7}{
\begin{tabular}{l|c|}
\cline{2-2}
                                                                                    & \textbf{Q3}             \\ \hline
\multicolumn{1}{|l|}{\textbf{Personal data about the driver}}                       & 69                      \\ \hline
\multicolumn{1}{|l|}{\textbf{Public data about the driver}}                         & 56                      \\ \hline
\multicolumn{1}{|l|}{\textbf{Public data not about the driver}}                     & 44                      \\ \hline
\multicolumn{1}{|l|}{\textbf{Special categories of personal data about the driver}} & \multicolumn{1}{l|}{15} \\ \hline
\multicolumn{1}{|l|}{\textbf{Financial data about the driver}}                      & \multicolumn{1}{l|}{15} \\ \hline
\multicolumn{1}{|l|}{\textbf{No data at all}}                                       & \multicolumn{1}{l|}{1}  \\ \hline
\end{tabular}}
}
\end{table}

To simplify interpretation and at the same time make it more expressive, answers were classified into the agreeing and disagreeing category, according to their level of agreement/disagreement. Participants that selected neither agree nor disagree are reported as undecided. 
All findings discussed below refer to the Table~\ref{tab:questions}, except question 3 which refers to Table~\ref{tab:question3}.
Question 2 has a high rate of agreement, so we can say that the participants agree that a modern car is similar to a modern computer.
From question 3 we note that the predominant categories according to respondents are: ``personal data about the driver'' (selected by 69\%); ``public data about the driver'' (selected by 56\%); ``public data not about the driver'' (selected by 44\%). Few participants think that their vehicle collects more sensitive data belonging to the special categories of personal and financial data (both 15\%). Just one participant thinks that modern cars do not collect any data at all.
Question 4 shows that 32\% of the participants agree with the statement above, moreover 35\% of them are undecided and 33\% of them disagree with the statement. It seems that participants are somehow equally distributed.
From the answers to question 5 it can be seen that just 26\% of the sample agrees to the transmission of data over the Internet, 26\% of participants are undecided while 48\% of them disagree with the statement. This means that the sample in general is not very convinced to send personal data over the Internet.
Question 6 shows that 80\% of the participants agree with the above statement and only 6\% disagree with the statement while 14\% of them are undecided.
The answers to question 7 show that a large part of the sample is undecided on this statement (41\%), 22\% of the participants agree with the statement while 37\% disagree.
Question 8 shows that  the 50\% of the participants agree with this statement while the 34\% are undecided and the rest of them (16\%) disagree.
Question 9 shows that 52\% of participants agree with this statement, 31\% of them are undecided and 17\% disagree with the statement.
The last question (Q10)  shows that just a minority of them (about 23\%) answered negatively to the question. The majority (51\%), agree that their data is protected using appropriate methodologies and techniques, indicating a good perception of security and trust on the part of drivers.

\begin{table}[ht]
\centering
\caption{Questionnaire responses in percent}
\label{tab:questions}
\renewcommand{\arraystretch}{1.3}
\scalebox{0.7}{
\begin{tabular}{c|c|c|c|c|c|c|c|c|}
\cline{2-9}
\multicolumn{1}{l|}{}                      & \textbf{Q2} & \textbf{Q4} & \textbf{Q5} & \textbf{Q6} & \textbf{Q7} & \textbf{Q8} & \textbf{Q9} & \textbf{Q10} \\ \hline
\multicolumn{1}{|c|}{\textbf{Agreeing}}    & 83\%        & 32\%        & 26\%        & 80\%        & 22\%        & 50\%        & 52\%        & 51\%         \\ \hline
\multicolumn{1}{|c|}{\textbf{Disagreeing}} & 6\%         & 33\%        & 48\%        & 6\%         & 41\%        & 34\%        & 31\%        & 23\%         \\ \hline
\multicolumn{1}{|c|}{\textbf{Undecided}}   & 11\%        & 35\%        & 26\%        & 14\%        & 37\%        & 16\%        & 17\%        & 26\%         \\ \hline
\end{tabular}}
\end{table}

In summary, the interviewed sample feel quite informed about modern vehicles. The majority of participants agrees that systems and technologies present in modern cars are increasingly similar to modern computers. Regarding the collection of personal data, the participants seem to be equally divided between those who agree, those who disagree and those who neither agree nor disagree. Moreover, according to the sample, the data collection is more oriented towards public and personal data, with no interest in financial information or special categories of personal data. The level of agreements regarding the collection of personal data may be due to the fact that drivers think it is neither necessary nor useful. The answers of question 9 tell us that half of the sample thinks that their data is analysed and studied by the vehicle systems in order to evaluate some personal aspects. This statement could have increased the level of disagreement of data collection, showing that there are some privacy concerns. Regarding the transmission of collected data just a few of the participants think it is truly necessary. This result could suggest that drivers have some privacy concerns about their personal data. In fact, considering the answers of question 5, almost half of the interviewed sample does not want personal data to be transmitted on the Internet by the vehicle. This kind of drivers may think that they have not enough control on their personal data once they are transmitted. Thus, it seems that drivers demonstrate to have some risk perceptions regarding their data. Recent attacks against car manufacturers that targeted drivers' personal data~\cite{cimpanu:buyer_data_breach,robinson:buyer_data_sold_bmw} could have influenced the drivers in this decision.

\section{Related work and conclusions}\label{sec:RWandConcl}
In 2014, Schoettle and Sivak~\cite{schoettle2014} surveyed public opinions in Australia, the United States and the United Kingdom regarding connected vehicles. Their research noted that people (drivers as well as non-drivers) expressed a high level of concern about the safety of connected cars, which does not seem surprising on the basis of the novelty of the concept at the time. 

In 2016, Derikx et al.~\cite{derikx2016} investigated whether drivers' privacy concerns can be compensated by offering monetary benefits. They analysed the case of usage-based auto insurance services where the rate is tailored to driving behaviour and measured mileage and found out that drivers were willing to give up their privacy when offered a small financial compensation.
We argue that the international research community may not have fully realised the necessity and relevance of such a compliance beyond its sheer legal urgency. This is confirmed by the scant literature focusing on protecting drivers' data. 
The ``CANDY'' attack reconfirms how data can be stolen following security weaknesses, which derive, in this particular case, from optimistic network isolation assumptions made at application layer~\cite{CANDY}.

A few works emphasise the overarching problem of how to effectively transmit the contents of a lengthy policy to people. Those wishing to use a service routinely accept the terms of the service provider without fully understanding them. As a result, users are not actually informed~\cite{pardo}. 
Furthermore, the literature referred to above reinforces the rationale for the data classification work carried out in this paper, because a comparative analysis of car manufacturers' privacy policies is not available. 

While there is some general awareness that treating people's personal data is essential to people's privacy and, consequently, freedom today, this paper showed that awareness to be very limited in the automotive domain. Car drivers' privacy concerns are lower than we think they should, especially given the quantity and quality of personal data that cars collect and manufacturers treat. This cannot be justified in terms of drivers' overall trust, which is found to be comparatively low. As a possible reason, privacy policies are found not to reach drivers well, indicating that the entire privacy area aboard modern cars demands immediate attention.

\section*{Acknowledgments}
This research has received funding from COSCA (COnceptualising Secure Cars)~\cite{COSCA}, a project supported by the European Union's Horizon 2020 research and innovation programme under the NGI TRUST grant agreement no 825618.

\bibliographystyle{ieeetr}
\bibliography{automotive}

\appendices
\section{Questionnaire}\label{app:questionnaire}

The core questionnaire questions are listed below.
\begin{itemize}
    \item[0.] How many hours a week do you drive a car?
    \begin{itemize}
        \item[$\circ$] 3-6 hours; $\circ$ 7-9 hour; $\circ$ 10-12 hours; $\circ$ 13-15 hours; $\circ$ 16-20 hours; $\circ$ 21+ hours
    \end{itemize}
\end{itemize}

\begin{enumerate}
    \item Are you knowledgeable about modern cars?
    \label{app:question1}
    \begin{itemize}
      \item[] Not at all $  \square-\square-\square-\square-\square-\square-\square$ Very knowledgeable about modern cars
    \end{itemize}

    \item How much do you agree with the following statement: a modern car is similar to a modern computer.
    \label{app:question2}
    \begin{itemize}
        \item[] Strongly disagree $\square-\square-\square-\square-\square-\square-\square$ Strongly agree
    \end{itemize}

    \item What kind of data do you think a modern car collects about its driver?
    \label{app:question3}

      \begin{itemize}

          \item[$\square$] No data at all;
          \item[$\square$] Public data not about the driver
          \item[$\square$] Public data about the driver
          \item[$\square$] Personal data about the driver (e.g. name, address, etc.)
          \item[$\square$] Special categories of personal data about the driver (e.g. racial or ethnic origin, political opinions, religious or philosophical beliefs,  biometric data, data concerning health or data concerning sex life or sexual orientation)
          \item[$\square$] Financial data about the driver (e.g. credit card number)

      \end{itemize}
    
    \item How much do you agree with the following statement: ``personal data collected by a modern car about its driver is necessary for the full functioning of the car".
    \label{app:question4}
    \begin{itemize}
        \item[] Strongly disagree $\square-\square-\square-\square-\square-\square-\square$ Strongly agree
    \end{itemize}

    \item How much do you agree with the following statement: ``it is necessary that the personal data collected by a modern car about its driver be transmitted over the Internet".
    \label{app:question5}
        \begin{itemize}
        \item[] Strongly disagree $\square-\square-\square-\square-\square-\square-\square$ Strongly agree
    \end{itemize}
    
    \item How much do you agree with the following statement: ``a modern car safeguards the life of its driver".
    \label{app:question6}
    \begin{itemize}
        \item[] Strongly disagree $\square-\square-\square-\square-\square-\square-\square$ Strongly agree
    \end{itemize}
    
    \item How much do you agree with the following statement: ``a modern car protects its driver's personal data better than it safeguards its driver's life".
    \label{app:question7}
    \begin{itemize}
        \item[] Strongly disagree $\square-\square-\square-\square-\square-\square-\square$ Strongly agree
    \end{itemize}

    \item How much do you agree with the following statement: ``a modern car processes the personal data it collects about its driver in a legitimate (i.e. coherently with pertinent regulations) way".
    \label{app:question8}
    \begin{itemize}
        \item[] Strongly disagree $\square-\square-\square-\square-\square-\square-\square$ Strongly agree
    \end{itemize}

    \item How much do you agree with the following statement: ``a modern car carries out a systematic and extensive evaluation of the personal data it collects about its driver on the basis of automated processing, including profiling (e.g. to  evaluate certain personal aspects of the driver to analyse or predict aspects concerning the driver's performance at work, economic situation, health, personal preferences, interests, reliability, behaviour, location or movements)".
    \label{app:question9}
    \begin{itemize}
        \item[] Strongly disagree $\square-\square-\square-\square-\square-\square-\square$ Strongly agree
    \end{itemize}
    
    \item How much do you agree with the following statement: ``the personal data a modern car collects about its driver is protected by suitable technology when the car transmits it somewhere on the Internet".
    \label{app:question10}
    \begin{itemize}
        \item[] Strongly disagree $\square-\square-\square-\square-\square-\square-\square$ Strongly agree
    \end{itemize}

\end{enumerate}

\end{document}